\documentclass[ reprint,nofootinbib,amsmath,amssymb,aps]{revtex4-1}
\usepackage{graphicx}
\usepackage{dcolumn}
\usepackage[colorlinks,linkcolor=magenta,anchorcolor=cyan,citecolor=blue]{hyperref}
\usepackage{bm}
\usepackage[multiple]{footmisc}
\newcommand{\fig}[1]{Fig.\ref{#1}}
\def\be{\begin{equation}}
\def\ee{\end{equation}}
\def\ba{\begin{eqnarray}}
\def\ea{\end{eqnarray}}

\def\nn{\nonumber}
\def\lf{\left}
\def\rt{\right}

\newcommand{\eq}[1]{(\ref{#1})}

\def\nn{\nonumber}\def\lf{\left}\def\rt{\right} \def\w{\omega}      \def\a {\alpha}  \def\d {\delta} \def\f {\phi} \def\g {\gamma} \def\h {\eta}  \def\k {\kappa} \def\l {\lambda} \def\z {\zeta} \def\x {\xi} \def\c {\chi}    \def\pd {\partial}   
\def\Q{\Theta} \def\W{\Omega} \def\Y {\Psi}    \def\S {\Sigma}  \def\F {\Phi}      \def\grad{\nabla}\def\.{\cdot}
\def\math {\mathcal}
\begin{document}

\title{Investigating two counting methods of the holographic complexity}
\author{Jie Jiang}
\email{jiejiang@mail.bnu.edu.cn}
\author{Bo-Xuan Ge}
\email{boxuange@mail.bnu.edu.cn}
\affiliation{Department of Physics, Beijing Normal University, Beijing 100875, China\label{addr1}}
\date{\today}

\begin{abstract}
We investigated the distinction between two kinds of ``Complexity equals Action"(CA) conjecture counting methods which are separately provided by Brown $ et\, al. $ and Lehner $et\, al.$ separately. For the late-time CA complexity growth rate, we show that the difference between two counting methods only comes from the boundary term of the segments on the horizon. However, both counting methods give the identical late-time result. Our proof is general, independent of the underlying theories of higher curvature gravity as well as the explicit stationary spacetime background. To be specific, we calculate the late-time action growth rate in SAdS black hole for F(Ricci) gravity, and show that these two methods actually give the same result. Moreover, by using the Iyer-Wald formalism, we find that the full action rate within the WDW patch can be expressed as some boundary integrations, and the final contribution only comes from the boundary on singularity.  Although the definitions of the mass of black hole has been modified in F(Ricci) gravity, its late-time result has the same form with that of SAdS black hole in Einstein gravity.
\end{abstract}
\maketitle

\section{Introduction}
Quantum computational complexity theory is used to figure out what the implications of quantum physics to computational complexity theory are \cite{1}. This theory never played a crucial role in gravitational physics until Susskind noticed one thing that there is a relationship between complexity and the deep inner structure of the stretched horizon as well as the extreme long-time behavior of black holes. Moreover, the complexity would determine what actions at one end of an Einstein-Rosen bridge can send signals to the other end. Furthermore, he argued that the difference between ``easy" operators and ``hard" operators is related to the time evolution of a certain measure of complexity which associated with the stretched horizon of Alice's black hole.\cite{2}. Whereafter, Susskind proposed the CV-conjecture and tried to build a bridge of complexity and the volume of the Einstein-Rosen bridge (ERB)\cite{3,4}. Roughly speaking, this duality can be described by
\ba
\math{C}=\frac{V}{G \ell_\text{AdS}}\,,
\ea
where $V$ is the volume of the ERB.

Although CV-duality has a lot of excellent properties, it also has some unsatisfactory features. For example, one has to choose a length scale \(\ell\). Meanwhile, there is a lack of a clear argument for the reason why the maximal volume slice should play the preferred role\cite{5,6}. Thus, soon afterward, Susskind proposed ``Complexity Equals Action" conjecture which is also known as CA-duality. This new ``bridge" connect the complexity with the action in bulk. Roughly speaking, this duality can be described by
\ba
\math{C}=\frac{I_{\text{WDW}}}{\pi \hbar}\,,
\ea
where $I_{\text{WDW}}$ is the action within the WDW patch.

The CA duality not only owns all good properties which the CV duality has but also gets rid of CA duality's drawbacks. So far, the CA-duality succeed in all the shock wave tests which the CV-duality passed before. And, this is the reason why the CA-duality been widely concerned. A lot of significant follow-up work has been made by a number of  brilliant physicist\cite{Jiang:1, Jiang:2, Jiang:3, Jiang:4, Jiang:5,Jiang:6,7,8,9,10,12,13,14,15,16,17,18,19,20,21,22,23,24,25, 26,27,28,29,31,32,33,34,35,36,37}.

There are two kinds of the typical method proposed separately by Brown $et.al$(BRSSZ)\cite{5,6} and Lehner $et.al$(LMPS)\cite{8}, to calculate the holographic complexity by CA duality. The BRSSZ method is put forward to calculating the complexity growth rate at the late time for CA duality. And the LMPS method is introduced to evaluate the complexity itself as well as its time evolution. In general, they are not equivalent. For instance, in the BRSSZ approach, there is no way to violate the Lloyd's bound,  while it is shown that the bound is violated by LMPS method due to the fact that the complexity growth approaches the bound at very late times from above. However, as illustrated by some explicit cases, these two methods will give the same CA complexity growth rate result at the very late time.

Therefore, in this paper, we focus on these two kinds of typical methods, and study the distinction between these two methods. We show that the difference between the two methods only comes from the boundary term of the segments near the horizon for the late-time action growth rate within the WDW patch. However, both counting methods give the identical late-time result.  After that, we consider Schwarzschild anti-de Sitter (SAdS) black hole in the F(Ricci) gravity and calculate its late-time action growth rate by using Iyer-Wald formalism\cite{38}. Then, we show the late-time limits of the action growth rate by these two methods actually give the same results.

The rest of this paper is structured as follows. In Sec.\ref{sec2}, we compare two methods firstly; then we show they give the same complexity growth rate result at late times. In Sec.\ref{sec3}, we briefly review the Iyer-Wald formalism in a general diffeomorphism covariant theory. Then, we calculate the action growth rate at late times for F$($Ricci$)$ gravity by using two methods separately. In the final Sec.\ref{sec4}, we conclude our paper with some discussions.

\section{late-time complexity growth rate}\label{sec2}

\subsection{Comparing two methods of CA duality}
\begin{figure}
\centering
\includegraphics[width=0.5\textwidth]{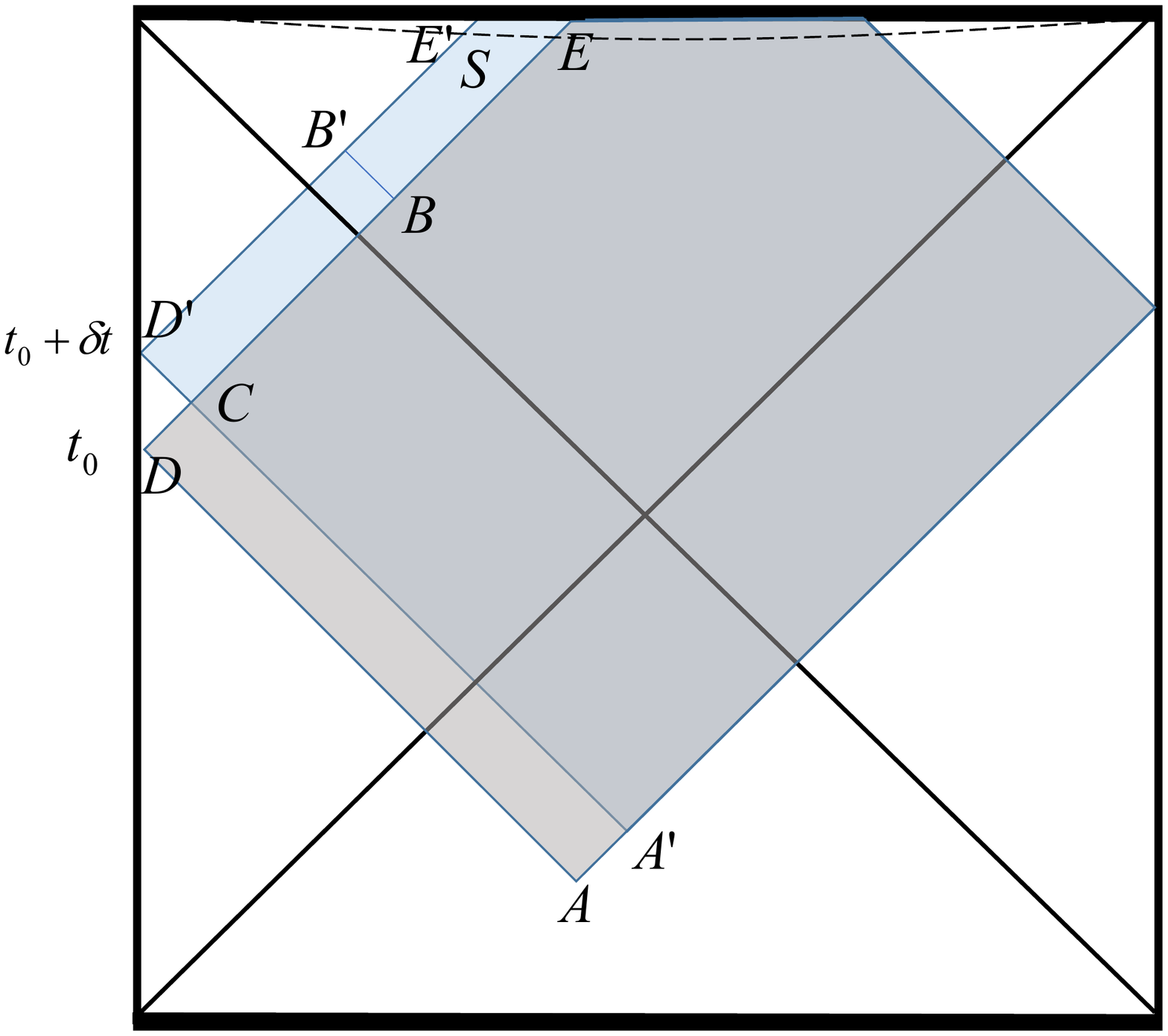}
\caption{Wheeler-DeWitt patch at the late time of spacetime which shares the same Penrose diagram with SAdS black hole, where the dashed line denotes the cut-off surface at the singularity, satisfying the asymptotic symmetries. }\label{WDW}
\end{figure}
As an example, firstly, we consider a stationary spacetime which shares the same Penrose diagram with the SAdS black hole. The Wheeler-DeWitt(WDW) patch of this spacetime can be illustrated in \fig{WDW}. For the corresponding action \( S(t_L,t_R)\), there are two choices of the time slice on the left and right boundaries. It is easy to figure out that the action is invariant under the time slices shifting. Namely, \(S(t_L+\delta t,t_R-\delta t)=S(t_L,t_R)\). However, we prefer to study some nontrivial cases. Thus, we vary the left asymptotic time left slice, meanwhile fix the time on the right boundary. Consequently, this can give us an opportunity to calculate the difference in the action between these two WDW patches.

The WDW patch which is corresponding to the \(t_0 \) is shown in dark color. \(S(t_0) \) denotes the action for this patch. Another patch corresponding to \(t_0+\d t\) is labeled by light color. Similarly, the action of this patch is denoted by \(S(t_0+\d t)\). Next, we will calculate the difference of the action between these two patches $\d S=S(t_0+\d t)-S(t_0)$, as illustrated in \fig{WDW} which can be obtained by
\ba\label{ag}
\d S=\d S'-\d S\,,
\ea
where $\d S',\d S$ denote the whole action (bulk action, surface term and boundary term) for the region $CD'E'E$ and $AA'CD$ separately.

According to the symmetries of this spacetime, one can find that geometric of the whole region $AA'CD$ is the same as the $BB'D'C$. Whence, the action growth \eq{ag} only comes from the total action of the region $BB'E'E$. At the late time, the segment $BB'$ will shrink to the horizon. Then, the segment $BE$ and $B'E'$ can be related by the time-translation symmetry. Thus, when we choose the parameter on the null segment invariant under the time evolution, it will not generate any effect on the total action growth. Finally, except for the bulk contribution from $BB'E'E$, there are only two surface contributions survival, one is from the surface $BB'$ near the horizon and the other one comes from $S=EE'$ near the singularity.

As suggested by Brown $et. al$ in Refs.\cite{5,6}, to evaluate the surface contribution from $BB'$ near the horizon, one can introduce a spacelike surface $\math{B}$, which is invariant under the time-transformation, to take place the segment $BB'$, and finally the surface contribution can be obtained by the limit $\math{B}\to \math{H}$.

On the other hand, Lehner $et.al$\cite{8} also introduce another counting method to evaluate this CA complexity. They use the whole action containing the null surface term and corner term to evaluate the action growth within the WDW patch straightly. Same analysis, the action growth also only comes from the region $BB'E'E$.

If we choose the affine parameter to the null generator of the null segment, which will leave no contribution from the null surface terms. We also make the null generator of the right future null segment invariant under the time transformation. Then, the part of action growth which comes from the corner terms of $E$ and $E'$, the counterterms from the null segment $BE$ and $B'E'$ will vanish. Moreover, since the horizon is generated by a killing vector, the counterterm from the null segment $BB'$ will also vanish.  Finally, the total contributions consist of the volume contribution from the areas $BB'E'E$, the surface contribution from the spacelike segment $S$, and the joint contributions from the $B$ and $B'$.

Lastly, the critical difference between the two methods is as follows: In the Brown $et.al$ calculation, the contribution which comes from the segment of the future horizon $BB'$ to \(\delta S\) does not vanish, where we use the limit of a spacelike segment to evaluate it. However this term plays no role in the calculation of Lehner $et.al$, and take place by two joints $B$ and $B'$ on the horizon, but this contribution is canceled in Brown $et.al$ method. Even so, in the next subsection, we will show that the final computation result is equal to a general higher curvature gravity.

One can verify that, for any stationary black hole, the difference of this two methods only come from the segment of the horizon for the calculation of the late-time complexity growth rate as shown in the last paragraph.

\subsection{Showing the equivalence for late-time action growth rate}
In this section, we show that the two methods will give the same complexity growth rate result at the late time. As shown in the last subsection, the difference between these two methods only comes from the boundary and joints terms of the segments on the horizon. In the Brown $et\,.al$ computation, the null boundary contribution is taken as the limit of the non-null general Gibbons-Hawking-York(GHY) surface term, i.e., we have
\ba\label{dI1}
\d I=\lim_{\math{B}\to\math{H}}4\int_{\math{B}}\Y^{ab}K_{ab}d\S\,,
\ea
where
\(\Y^{ab}=E_R^{acbd}n_cn_d\)
is equiped with
\ba
E_R^{abcd}=\frac{\pd \math{L}_\text{grav}}{\pd R_{abcd}}\,
\ea
and the normal $1$-form $n_a$ of the segment $\math{B}$,  $K_{ab}=h_a{}^c\grad_c n_b$ is the extrinsic curvature of this surface, \(d\Sigma\) is the volume element on the segment $\math{B}$.

Since the spacetime is well defined near the horizon and $E_R^{abcd}$ is constructed by $g_{ab}$ and $R_{abcd}$, then, $E_R^{abcd}$ will be well defined as well. Considering two arbitrary vectors $u^a$ and $v^a$ with finite length, since $n_{[a}u_{b]}$ and $n_{[a}v_{b]}$ are in proportion to the bi-normal on the subspace span$\{n_a,u_a\}$ and span$\{n_a,v_a\}$ separately. Then, $n_{[a}u_{b]}$ and $n_{[a}v_{b]}$ will be well defined, which implies that
\ba
E_R^{acbd}n_an_cu_cv_d
\ea
will be a finite quantity near the horizon.

For simplicity, we assume that the segment $\math{B}$ can be described by the scalar function $\c=\sqrt{\x^a \x_a}$ with the killing vector $\x^a$ on the Killing horizon $\math{H}$. The Killing horizon can be obtained by setting \(\chi =0\). Note that $\c$ is an infinitesimal parameter near the horizon. The outer directed  normal vector of the segment $\math{B}$ can be written as
\ba\label{na}
n_a=-\a \grad_a \c =  -\a \c^{-1}\x^b\grad_a\x_b\,,
\ea
Since $\math{L}_{\x}g_{ab}=2\grad_{(a}\x_{b)}=0$, one can obtain
\ba
\x^a n_a=-\a\c^{-1}\x^{(a}\x^{b)}\grad_{(a}\x_{b)}=0\,.
\ea
That is to say, $\x^a$ is the tangent vector on the segment $\math{B}$, and as long as $\math{B}$ is non-null surface, $\x^a$ and $n^a$ are  linear independence. One can define the normal vector
\ba
r^a=\c^{-1} \x^a\,.
\ea
Then, for any point on the non-null segment $\math{B}$, the complete relation can be shown as
\ba
\d_a{}^b=-n_an^b+r_a r^b+\g_a{}^b\,.
\ea
Where \(\g_a{}^b\) is independent with \(n_an^b\) and \(r_a r^b\). Using this complete relation, we have
\ba\label{YK}\begin{aligned}
\Y^{ab}K_{ab}&=\Y^{cd}r_cr_dr^ar^b\grad_an_b+2\Y^{cd}r_cr^a\g^b{}_d\grad_an_b\,.\\
&+\Y^{cd}\g^a{}_c\g^b{}_d\grad_an_b
\end{aligned}\ea
For the first term, in virtue of  $\math{L}_{\x}d\c=d\math{L}_{\x}\c=0$, one can find $\math{L}_\x n^b=\x^a \grad_an^b-n^a\,. \grad_a\x^b=0$.

Then the first term of Equ.(9) can be written as
\ba\begin{aligned}\label{di1}
4\Y^{cd}r_cr_dr^ar^b\grad_an_b&=\c^{-2}\hat{\Y}\x^a\x^b\grad_an_b\\
&=\c^{-2}\hat{\Y}n^a\x^b\grad_a\x_b\\
&=-\a^{-1}\c^{-1}\hat{\Y}n^an_a\\
&=\a^{-1}\c^{-1}\hat{\Y}\,,
\end{aligned}\ea
where \(\hat{\Y}=4\Y^{ab}r_a r_b.\) To obtain the value of Equ.(9) on the horizon, we like to introduce a special coordinate. On the segment $\math{B}$($\c_0$), one can choose the coordinate $\{\l,y^A,A=1,\cdots,n-2\}$ such that $\lf(\frac{\pd}{\pd \l}\rt)^a=\x^a$ is a Killing field on the horizon. Then, the induce metric on this segment can be described as
\ba\label{spline}
d\hat{s}^2=\c^2 d\l^2+2N_A(\c,y) d\l dy^A+\g_{AB}(\c,y)dy^Ady^B\,.\nn\\
\ea
Note $\f(p,\h)$ as the diffeomorphism of the vector field $n^a$. For any $q=\f(p,\h)\in \math{B}(\c)$ and $ p \in \math{B}(\c_0) $, we can define its coordinate as $(\c,\l_p,y^A_p)$. Using the relation $n^a n_a=-1$, one can find
\ba
n^a=\a(\c,y)^{-1}\lf(\frac{\pd}{\pd \c}\rt)^a\,.
\ea
Then, the line element can be shown as
\ba\begin{aligned}
ds^2&=-\a(\c,y)^2d\c^2+\c^2 d\l^2+2N_A(\c,y) d\l dy^A\\
&+\g_{AB}(\c,y)dy^Ady^B\,.
\end{aligned}\ea
According to the induced spatial line element \eq{spline}, one can obtain
\ba\label{dS}
d\S=\sqrt{\c^2-N_A N^A}\sqrt{\g}d\l d^{n-2}y\,.
\ea
On the horizon, $\x^a$ becomes the normal vector, $i.e.$, we have
\ba
\lim_{\math{B}\to \math{H}}g_{ab} \lf(\frac{\pd}{\pd \l}\rt)^a \lf(\frac{\pd}{\pd y^A}\rt)^b=\lim_{\c\to 0} N_A(\c,y)=0\,.
\ea
That is to say, $N_A$ is an infinitesimal parameter when $\c\to 0$. So we can set
\ba\label{NAj}
N_A\propto \c^{j_A}\,,\ \ \ \ j_A>0
\ea
under the leading order approximation of $\c$. One can verify that if $j_A\leq 1$, the Ricci-scalar function will diverge when $\c\to 0$. However, in this paper, $\c=0$ denotes the killing horizon where the curvature is well defined. That is to say, to ensure the curvature finite on the Killing horizon, we must have $j_A>1$. Then, under the leading order
approximation of $\c$, \eq{dS} becomes
\ba
d\S=\c\sqrt{\g}d\l d^{n-2}y\,.
\ea
The extrinsic curvature can be written as
\ba\label{Kab}
K_{ab}=\frac{1}{2}\math{L}_{n}h_{ab}=\frac{1}{2\a} \frac{\pd h_{ab}}{\pd \c}\,.
\ea
Using the relation $\k^2=\lim_{\c\to0}\lf(\grad_a \x_b \grad^a \x^b\rt)$ with the surface gravity $\k$ on the horizon, one can obtain
\ba\label{ak}
\a=-\k^{-1}\,.
\ea
According to Eqs. \eq{di1}, \eq{dS} and \eq{ak}, the first term contribution in \eq{dI1} can be shown as
\ba
\d I_1=-\int_{\math{H}}\k \hat{\Y}\sqrt{\g}d\l d^{n-2}y\,.
\ea
For the second term contribution
\ba\label{dI2}\begin{aligned}
\d I_2&= 8\lim_{\c\to 0} \int_{\math{B}}\Y^{cd}r_cr^a\g^b{}_dK_{ab}\\
&=8\lim_{\c\to 0} \int_{\math{B}}\Y^{cA}r_c K_{\l A}\sqrt{\g}d\l d^{n-2}y\,.
\end{aligned}\ea
According to \eq{Kab}, one can obtain $K_{\l A}=\frac{\pd_\c N_A}{2\a}$. Using \eq{NAj}, \eq{dI2} becomes
\ba
\d I_2=0\,.
\ea
 Similarly, the last term contribution is
\ba\begin{aligned}
\d I_3&=4\lim_{\c\to 0} \int_{\math{B}}\Y^{cd}\g^a{}_c\g^b{}_d K_{ab}d\S\\
&=4\lim_{\c\to 0} \int_{\math{B}}\Y^{AB}K_{AB}\c\sqrt{\g} d\l d^{n-2}y\\
&=4\lim_{\c\to 0} \int_{\math{B}}\Y^{AB}\frac{\pd_\c \g_{AB}}{2\a}\c\sqrt{\g} d\l d^{n-2}y\,.
\end{aligned}\ea
If we assume that this term is non-vanish, i.e., under the leading order of $\c$, we have
\ba
\pd_\c\g_{AB}\propto \c^{l}\,,\ \ \ \ \ l\leq-1\,,
\ea
then, $\g_{AB}\propto \c^{l+1}$ when $l<-1$ or $\g_{AB}\propto \ln\c$ when $l=-1$ will diverge. That is to say, this contribution also vanishes, i.e., $\d I_3=0$.

Using above result, we have
\ba\label{dIf}
\d I=-\int_{\math{H}}\hat{\Y}\k \sqrt{\g} d\l d^{n-2}y\,,
\ea
which is actually the surface term of the null boundary. In \cite{Jiang:2}, we have obtained the full boundary term which contain the boundary term as well as the counterterm,
\ba\begin{aligned}
\d I_\text{bdry}=&-\int_{\math{B}}\hat{\Y}\k \sqrt{\g} d\l d^{n-2}y-\int_{\math{B}}\hat{\Q}\ln\lf(l_\text{ct}\Q\rt) \sqrt{\g} d\l d^{n-2}y\\
&-\int_{\math{C}_1}\hat{\Y}\h \sqrt{\g} d^{n-2}y+\int_{\math{C}_2}\hat{\Y}\h \sqrt{\g} d^{n-2}y\,,
\end{aligned}\ea
where $\hat{\Q}=\grad_a(k^a\hat{\Y})=\frac{1}{\sqrt{\g}}\pd_\l (\hat{\Y}\sqrt{\g})$ and $\hat{\Q}=\grad_a{k^a}=\frac{1}{\sqrt{\g}}\pd_\l (\sqrt{\g})$ are the expansion scalar of the null generator with an arbitrary length scale $l_\text{ct}$, $\h=\ln\lf(-\frac{1}{2}k\.l\rt)$ is the transformation parameter of the corner. This full boundary action is invariant under the reparameterization of the null generator $k^a$ of the null segment.

For investigating the action growth rate within WDW patch at a late time,  the counterterm contribution from the null boundary which is a segment of the horizon will vanish. We choose $l^a$ as the affinely null generator of the null boundary $\math{N}$ satisfying $\math{L}_{\x}l^a=0$. Then, the time derivative of this surface term and counterterms will vanish. If we choose $\x^a$ as the null generator of the null segment $\math{B}$, since $\math{L}_{\x}\lf(l^a\x_a\rt)=0$, i.e., $\math{L}_{\x}\h=0$, the corner contribution from $\math{C}$ also vanishes. Finally, the action growth rate of the full action only comes from the null surface term on the horizon, i.e.,
\ba\begin{aligned}
\d I_\text{bdry}&=-\int_{\math{H}}\hat{\Y}\k \sqrt{\g} d\l d^{n-2}y\\
&=-\d t\int_{\math{H}}\hat{\Y}\k \sqrt{\g}  d^{n-2}y\,,
\end{aligned}\ea
 which is obviously the result obtained from the BRSSZ method. On the other hand, since the full action is invariant under the reparameterization of the null generator. We can also choose the affine parameter to the null generator $k^a$ of the null segment $\math{B}$, then, this null surface term will vanish. The non-vanish contribution only comes from the corner term $\math{C}$, which is actually the progress proposed by Lehner $et.al$. This affine null generator $k^a$ can be written as $k^a=e^{-\k \l}\x^a$. The transformation parameter can be shown as
\ba
\h(\l)=\ln\lf(-\frac{1}{2}k\.l\rt)=-\k\l+\ln\lf(-\frac{1}{2}\x\.l\rt)\,,
\ea
then, we have
\ba\begin{aligned}
\d I_\text{bdry}&=-\int_{\math{C}_1}\hat{\Y}\h \sqrt{\g} d^{n-2}y+\int_{\math{C}_2}\hat{\Y}\h \sqrt{\g} d^{n-2}y\\
&=-\d t\int_{\math{H}}\hat{\Y}\k \sqrt{\g}  d^{n-2}y\,,
\end{aligned}\ea
which gives the same result as \eq{dIf} by the BRSSZ method.

Moreover, this result reveals that the difference between these two methods only comes from the boundary term of the segments on the horizon for the calculation of late-time rate. However, both counting methods give the identical result.

\section{The late-time action growth rate for F(Ricci) gravity}\label{sec3}

\subsection{Iyer-Wald formalism}

 We consider a diffeomorphism covariant theory on an $n$-dimensional oriented manifold $\mathcal{M}$, where the Lagrangian $n$-form $\bm{L}=L\bm{\epsilon}$ is supposed to be constructed locally out of the metric $g_{ab}$, other matter fields $\psi$, as well as the symmetrized covariant derivatives of the corresponding Riemann tensor $R_{abcd}$ and $\psi$, with $\epsilon$ the volume element compatible with the metric on the manifold $\mathcal{M}$\cite{38}. We use $\phi=(g_{ab},\psi)$ to denote all dynamical fields and perform a variation of $\bm{L}$, which leads to
 	\begin{eqnarray}
 		\delta\bm{L}=\bm{E}\delta\phi + d\bm{\Theta}(\phi,\delta\phi),\label{1}
 	\end{eqnarray}
 	where
  $\bm{E}=0$ correspond to the equations of motion of the theory, and $\bm{\Theta}$  is called the symplectic potential $(n-1)$-form. The symplectic current $(n-1)$-form $\omega$ is then defined by
 	\begin{eqnarray}
 		\bm{\omega}(\phi,\delta_{1}\phi,\delta_{2}\phi)=\delta_{1}\bm{\Theta}(\phi,\delta_{2}\phi)-\delta_{2}\bm{\Theta}(\phi,\delta_{1}\phi),\label{2}
 	\end{eqnarray}
 	where $\delta_{1}$ and $\delta_{2}$ denote the variations with respect to different parameters.
 	
 	The Noether current $(n-1)$-form $\bm{J}_{\z}$ associated with an arbitrary smooth vector field $\z^a$ is defined as
 	\begin{eqnarray}\label{J1}
 		\bm{J}_{\z}=\bm{\Theta}(\phi,\mathcal{L}_{\z}\phi)-\z\cdot\bm{L},\label{3}
 	\end{eqnarray}
 	where we replace $\delta$ by $\mathcal{L}_{\z}$ in the expression of $\bm{\Theta}$ and the `dot' represents the contraction of $\z^a$ into the first index of $\bm{L}$. A simple calculation gives
 	\begin{eqnarray}
 		d\bm{J}_{\z}=-\bm{E}\mathcal{L}_{\z}\phi,\label{4}
 	\end{eqnarray}
 	which indicates $d\bm{J}_{\z}=0$ when the equations of motion are satisfied. On the other hand, as shown in \cite{39}, the Noether current $(n-1)$-form can also be expressed in the following form
 	\begin{eqnarray}
 		\bm{J}_{\z}=\bm{C}_{\z}+d\bm{Q}_{\z}.\label{5}
 	\end{eqnarray}
 Here $\bm{Q}_{\z}$ is the so-called Noether charge associated with $\z^a$ and $\bm{C}_{\z}=\z \cdot \bm{C}$ are interpreted as the corresponding constraints of the theory, which vanish when the equations of motion are satisfied. As shown in \cite{38}, this $(n-2)$-form can always be expressed as
\ba\label{K}
\bm{Q}_\z=\bm{W}_c\z^c+\bm{X}^{cd}\grad_{c}\z_{d}\,,
\ea
where
\ba\label{XW}\begin{aligned}
\lf(\bm{X}^{cd}\rt)_{c_3\cdots c_n}&=-E_R^{abcd}\bm{\epsilon}_{abc_3\cdots c_n}\,,\\
\lf(\bm{W}^d\rt)_{c_3\cdots c_n}&=2\grad_c E_R^{abcd}\bm{\epsilon}_{abc_3\cdots c_n}\,.
\end{aligned}\ea

\subsection{Late-time action growth rate}
Considering the $F($Ricci$)$ gravity, its corresponding bulk Lagrangian form can be written as $$\bm{L}=F(R_{ab})\bm{\epsilon}\,.$$
In this paper, we only focus on a particular case of this theory, where the SAdS black hole can be regarded as its solution. The corresponding line element is given by
\ba\label{ds2}
ds^2=-f(r)dt^2+\frac{dr^2}{f(r)}+r^2 d\W^2_{k,n-2}
\ea
where $f(r)=\frac{r^2}{L^2}+k-\frac{\w^{n-3}}{r^{n-3}}$
is the blackening factor, with the AdS curvature radius $L$. $k=\{+1,0,-1\}$ denotes the $(n-2)$-dimensional spherical, planar, and hyperbolic geometry individually. According to this metric, one can find
\ba\label{Rab}
R_{ab}=\frac{n-1}{L^2}g_{ab}\,.
\ea
Then, we can further obtain
\ba
\bm{W}^a=0
\ea
and $\hat{\Y}$ is a constant. For the action growth within the WDW patch, the bulk contribution only comes from the bulk region $M=BB'E'E$. At the late time, it can be generated by the killing vector $\x^a=\lf(\frac{\pd}{\pd t}\rt)^a$ through the null hypersurface $\math{N}=BE$. Then, we have
\ba
\d I_\text{bulk}=\int_{ M}\bm{L}=\d t\int_{\math{N}}\x\.\bm{L}\,.
\ea
According to \eq{J1} and the relation $\bm{\Q}\left(g_{ab},\math{L}_{\x}g_{ab}\rt)=0$, we have
\ba
\d I_\text{bulk}=-\d t\int_{\math{N}}d \bm{Q}_\x=-\d t\int_{B} \bm{Q}_\x-\d t\int_{E} \bm{Q}_\x\,.
\ea
By using the relation $\grad_a\x_b=\k \bm{\epsilon}_{ab}$ with $\k=f'(r_h)/2$ on the horizon, the first term can be written as
\ba
-\d t\int_{B} \bm{Q}=\d t\int_{\math{H}}\hat{\Y}\k \sqrt{\g}  d^{n-2}y=\d t\W_{k,n-2}\F_h\k\,.
\ea
where we denote $\F(r)=\hat{\Y} r^{n-2}$ and $\F_h=\F(r_h)$. For the second term, we have
\ba\begin{aligned}
-\d t\int_{E} \bm{Q}&=- \d t\int_{\S_0}E_R^{abcd}\grad_{c}\x_{d}\bm{\epsilon}_{ab}r^{n-2}d\S_{k,n-2}\\
&=-\frac{1}{2} \d t\W_{k,n-2}\lim_{r\to0}\lf[r^{n-2}f'(r)\hat{\Y}\rt]\\
&=-\frac{(n-3)}{2}\hat{\Y}\W_{k,n-2}\w^{n-3} \d t\,,
\end{aligned}\ea
where we used the relation $\grad_a\x_b=-\frac{1}{2}f'(r)\bm{\epsilon}_{ab}$ and $\hat{\Y}$ is a constant on the horizon. According to
\eq{Rab}, one can find that $\Y_{ab}=\frac{1}{4}\hat{\Y}h_{ab}$. Then, the surface term near the singularity becomes
\ba\begin{aligned}
I_{\math{S}}&=4\int_{\math{S}}\Y^{ab}K_{ab}d\S\\
&=\hat{\Y}\int_{\math{S}}K d\S\\
&=-\frac{(n-1)}{2}\hat{\Y}\W_{k,n-2}\w^{n-3}\d t\,,
\end{aligned}\ea
where we have used the expression
\ba
K=-\frac{1}{r^{n-2}\sqrt{-f}}\frac{d}{dr}\lf(r^{n-2}\sqrt{-f}\rt)
\ea
for the spacelike surface $E$ $(r=r_\epsilon)$ and let $r_\epsilon\to0$ in the end.

Finally, we turn to evaluate the contribution from the segment of the horizon. By using the BRSSZ method, according to \eq{dI1}, we have
\ba\begin{aligned}\label{horizon}
I_\math{H}&=4\lim_{\math{B}\to \math{H}}\int_{\math{B}}\Y^{ab}K_{ab}d\S\\
&=\d t\W_{k,n-2}\lim_{r\to r_h}\lf(r^{n-2}\sqrt{-f}\hat{\Y}K\rt)\\
&=-\d t\W_{k,n-2}\F_h\k\,.
\end{aligned}\ea
And we can see that this term will counteract the bulk boundary term from the null joint $B$ on the horizon.

Next, we use the LMPS method to evaluate this contribution. If we choose the affine parameter to the null generator of the null segments. Then, the non-vanish contributions only come from the joints $B$ and $B'$. And the null generators for the null segment $DE$ and $AD$ can be chosen as $k_{1a}=\grad_a u$ and $k_{2a}=-\grad_a v$ separately. Here the null coordinates are defined as $u=t+r^*(r)$ and $v=t-r^*(r)$ with $r^*(r)=\int f^{-1}dr$. Then, we have $k_1\cdot k_2=-2/f$. The joint contribution becomes
\ba\begin{aligned}
&\d I_\text{joint}=I_{B'}-I_{B}\\
&=\W_{k,n-2}\d r \lf.\frac{\pd}{\pd r}\lf[\F(r)\ln \lf(-f(r)\rt)\rt]\rt|_{r=r_B}\\
&=-\frac{\W_{k,n-2}\d t}{2} \lf[\F(r_B)f'(r_B)+\F'(r_B)f(r_B)\ln \lf(-f(r_B)\rt)\rt]\,,\\
\end{aligned}\ea
where we have used
\ba
\d r=r_B'-r_B=-\frac{1}{2}f(r_B)\d t\,.
\ea
At the late time, since $r_B\to r_h$, we have
\ba
\d I_B=-\d t \W_{k,n-2}\F_h\k\,,
\ea
which exactly gives the same result as Eq.\eq{horizon} by the BRSSZ method. Combining these results, the action growth rate can be written as
\ba
\frac{d I}{dt}=-(n-2)\hat{\Y}\W_{k,n-2}\w^{n-3}=2 M\,,
\ea
where
\ba
M=-\frac{(n-2)}{2}\hat{\Y}\W_{k,n-2}\w^{n-3}
\ea
is the ADT mass of the SAdS black hole in F(Ricci) gravity.

\section{Conclusion} \label{sec4}
There are a lot of following works are proposed after CA duality. It contains calculation methods and correction terms. There are two kinds of the typical method proposed separately by Brown $et.al$(BRSSZ)\cite{5,6} and Lehner $et.al$(LMPS)\cite{8}, to calculate the holographic complexity with CA duality. In this paper, we showed the differences between BRSSZ and LMPS methods of late-time complexity growth rate, which come from the boundary term of the segments on the horizon. And both counting methods give an identical result of the late-time rate for a general higher curvature gravity although they are not equivalent for complexity itself. Our proof is universal, independent of the underlying theories of higher curvature gravity as well as the explicit stationary spacetime background. To be specific, we calculated the action growth rate for F(Ricci) gravity using these two methods and showed their equivalency at the late-time complexity calculation. Using the Iyer-Wald formalism, we show that the bulk contribution can be written as two boundary contributions connected to the Noether charge. And the late-time rate only comes from the contributions from the singularity for the SAdS black hole. Moreover, the result of SAdS black hole in F(Ricci) gravity is consistent with the result in Einstein gravity although the mass is modified in F(Ricci) gravity.

\section*{Acknowledge}
J.J. is partially supported by NSFC with Grant No.11375026, 11675015, and 11775022. B.X.G. is supported in part by FWO-Vlaanderen through the project G020714N, G044016N, and G006918N. He is also an individual FWO Fellow supported by 12G3515N. This research was supported by NSFC Grants No. 11775022 and 11375026.
\\
\\
\\
\\


\begin{thebibliography}{100}
\bibitem{1} John Watrous, ``Quantum Computational Complexity," arXiv:0804.3401.
\bibitem{2} L. Susskind, ``Butterflies on the Stretched Horizon," arXiv:1311.7379.
\bibitem{3} D. Stanford and L. Susskind, ``Complexity and shock wave geometries," Phys. Rev. D {\bf90} 126007 (2014).
\bibitem{4} Leonard Susskind, Ying Zhao,``Switchbacks and the Bridge to Nowhere," arXiv:1408.2823
\bibitem{5} Adam R. Brown, Daniel A. Roberts, Leonard Susskind, Brian Swingle, Ying Zhao,``Complexity, action, and black holes,"Phys. Rev. D {\bf93}, 086006 (2016).
\bibitem{6} Adam R. Brown, Daniel A. Roberts, Leonard Susskind, Brian Swingle, Ying Zhao,``Holographic Complexity Equals Bulk Action?" Phys. Rev. Lett. {\bf116}, 191301 (2016)
\bibitem{7} Leonard Susskind,``Black Holes and Complexity Classes," arXiv:1802.02175
\bibitem{8} L. Lehner, R. C. Myers, E. Poisson and R. D. Sorkin, ``Gravitational action with null boundaries," Phys. Rev. D {\bf94}, 084046 (2016).
\bibitem{9} R. A. Jefferson and R. C. Myers, ``Circuit complexity in quantum field theory," JHEP {\bf1710} 107 (2017).
\bibitem{10} K. Hashimoto, N. Iizuka and S. Sugishita, ``Time evolution of complexity in Abelian gauge theories," Phys. Rev. D {\bf96}  126001 (2017).
\bibitem{12} A. Reynolds and S. F. Ross,``Complexity in de Sitter Space," Class. Quant. Grav. {\bf34},175013 (2017).
\bibitem{13} S. Chapman, H. Marrochio and R. C. Myers, ``Complexity of Formation in Holography," JHEP {\bf1701}  062 (2017).
\bibitem{14} D. Carmi, R. C. Myers and P. Rath, ``Comments on Holographic Complexity," JHEP {\bf1703} 118 (2017).
\bibitem{15} B. Czech, ``Einstein Equations from Varying Complexity,'' Phys. Rev. Lett. {\bf120} 031601 (2018).
\bibitem{16} P. Caputa, N. Kundu, M. Miyaji, T. Takayanagi and K. Watanabe, ``Liouville Action as Path-Integral Complexity: From Continuous Tensor Networks to AdS/CFT,'' JHEP {\bf 1711} 097 (2017)
\bibitem{17} M. Alishahiha, ``Holographic Complexity,'' Phys. Rev. D {\bf92}  126009 (2015).
\bibitem{18} P. Caputa, N. Kundu, M. Miyaji, T. Takayanagi and K. Watanabe, ``Anti-de Sitter Space from Optimization of Path Integrals in Conformal Field Theories,'' Phys. Rev. Lett. {\bf119}, 071602 (2017).
\bibitem{19} A. R. Brown and L. Susskind, ``Second law of quantum complexity,'' Phys. Rev.  D {\bf 97} 086015 (2018).
\bibitem{20}  C.~A.~Agón, M.~Headrick and B.~Swingle, ``Subsystem Complexity and Holography,''  JHEP {\bf 1902}, 145 (2019).
\bibitem{21} O. Ben-Ami and D. Carmi, ``On Volumes of Subregions in Holography and Complexity,'' JHEP {\bf1611}, 129 (2016).
\bibitem{22} S. Chapman, M. P. Heller, H. Marrochio and F. Pastawski, Phys. Rev. Lett. {\bf120} 121602 (2018).
\bibitem{23} Y. Zhao,  Phys. Rev. D {\bf 97} 126007 (2018).
\bibitem{24} Z. Fu, A. Maloney, D. Marolf, H. Maxfield and Z. Wang, JHEP {\bf02} 072 (2018).
\bibitem{25}  L.~Hackl and R.~C.~Myers, ``Circuit complexity for free fermions,'' JHEP {\bf 1807}, 139 (2018).
\bibitem{26}  M.~Alishahiha, A.~Faraji Astaneh, M.~R.~Mohammadi Mozaffar and A.~Mollabashi, ``Complexity Growth with Lifshitz Scaling and Hyperscaling Violation,'' JHEP {\bf 1807}, 042 (2018)
\bibitem{27} J. Couch, S. Eccles, W. Fischler and M. L. Xiao, ``On Volumes of Subregions in Holography and Complexity," JHEP {\bf1803} 108 (2018).
\bibitem{Jiang:1} J.~Jiang and X.~Liu,``Circuit Complexity for Fermionic Thermofield Double states,'' Phys.\ Rev.\ D {\bf 99} 026011 (2019).
\bibitem{28} B.~Swingle and Y.~Wang, ``Holographic Complexity of Einstein-Maxwell-Dilaton Gravity,'' JHEP {\bf 1809}, 106 (2018).
\bibitem{29} M. Moosa, ``Evolution of Complexity Following a Global Quench," JHEP {\bf1803} 031(2018).
\bibitem{Jiang:2}
 J.~Jiang and H.~Zhang, ``Surface term, corner term, and action growth in F(Riemann) gravity theory,'' Phys.\ Rev.\ D {\bf 99}, no. 8, 086005 (2019).
\bibitem{Jiang:3}  J.~Jiang, ``Holographic complexity in charged Vaidya black hole,'' Eur.\ Phys.\ J.\ C {\bf 79}, 130 (2019).
\bibitem{Jiang:4}  J.~Jiang, ``Action growth rate for a higher curvature gravitational theory,'' Phys.\ Rev.\ D {\bf 98} 086018 (2018).
\bibitem{Jiang:5}  J.~Jiang, J.~Shan and J.~Yang,``Circuit complexity for free Fermion with a mass quench,'' arXiv:1810.00537.
\bibitem{Jiang:6}  J.~Jiang and X.~W.~Li, ``Modified "complexity equals action" conjecture,'' arXiv:1903.05476.
\bibitem{31} D. Carmi, R. C. Myers, and P. Rath, ``Comments on Holographic Complexity'' JHEP {\bf03} 118(2017).
\bibitem{32} L. Lehner, R. C. Myers, E. Poisson, and R. D. Sorkin, ``Gravitational action with null boundaries,'' Phys. Rev. D {\bf94} 084046, (2016).
\bibitem{33} D. Carmi, S. Chapman, H. Marrochio, R. C. Myers and S. Sugishita, ``On the Time Dependence of Holographic Complexity,' JHEP {\bf1711} 188 (2017).
\bibitem{34} S. Chapman, H. Marrochio and R. C. Myers, ``Holographic complexity in Vaidya spacetimes. Part I,'' JHEP {\bf 1806} 046(2018).
\bibitem{35}  S.~Chapman, H.~Marrochio and R.~C.~Myers, ``Holographic complexity in Vaidya spacetimes. Part II,'' JHEP {\bf 1806}, 114 (2018).
\bibitem{36}  B.~Chen, W.~M.~Li, R.~Q.~Yang, C.~Y.~Zhang and S.~J.~Zhang, ``Holographic subregion complexity under a thermal quench,''
  JHEP {\bf 1807}, 034 (2018).
\bibitem{37} R. G. Cai, M. Sasaki and S. J. Wang, ``Action growth of charged black holes with a single horizon,'' Phys.\ Rev.\ D {\bf 95}, 124002 (2017).
\bibitem{38} V. Iyer and R. M. Wald, ``Some properties of Noether charge and a proposal for dynamical black hole entropy,'' Phys. Rev. D {\bf 50}, 864 (1994).
\bibitem{39} V. Iyer and R. M. Wald, ``Comparison of the Noether charge and Euclidean methods for computing the entropy of stationary black hole,'' Phys. Rev. D {\bf 52}, 4430 (1995).


\end{thebibliography}
\end{document}